\documentclass[12pt]{article}
  \usepackage{amsfonts}
  \usepackage{amsmath}
\usepackage{amssymb}
\usepackage{amscd}
\usepackage[dvips]{graphicx}
\begin{document}
~~
\bigskip
\bigskip
\begin{center}
{\Large {\bf{{{Twisted acceleration-enlarged Newton-Hooke Hopf
algebras}}}}}
\end{center}
\bigskip
\bigskip
\bigskip
\begin{center}
{{\large ${\rm {Marcin\;Daszkiewicz}}$ }}
\end{center}
\bigskip
\begin{center}
{ {{{Institute of Theoretical Physics\\ University of Wroc{\l}aw pl.
Maxa Borna 9, 50-206 Wroc{\l}aw, Poland\\ e-mail:
marcin@ift.uni.wroc.pl}}}}
\end{center}
\bigskip
\bigskip
\bigskip
\bigskip
\bigskip
\bigskip
\bigskip
\bigskip
\begin{abstract}
Ten Abelian twist deformations of acceleration-enlarged Newton-Hooke
Hopf algebra are considered. The  corresponding quantum space-times
are derived as well. It is demonstrate that their contraction limit
$\tau \to \infty$ leads to the new  twisted acceleration-enlarged
Galilei spaces.
\end{abstract}
\bigskip
\bigskip
\bigskip
\bigskip
\bigskip
\bigskip
\bigskip
\bigskip
\bigskip
 \eject
\section{{{Introduction}}}

The two  Newton-Hooke cosmological algebras ${ NH}_{\pm}$ were
introduced in the framework of  classification of all kinematical
groups \cite{bacry}. Both  algebras contain the characteristic
cosmological time scale $\tau$  which can be interpreted in terms of
the inverse of Hubble's constant
 for the expanding universe $({ NH}_{+})$ or associated
with  the "period" for the oscillating case $({ NH}_{-})$. For time
parameter $\tau$ approaching infinity we  get the Galilei group
$\,{\mathcal G}$ acting on the standard (flat) nonrelativistic
space-time.

Recently, there were proposed two acceleration-enlarged Newton-Hooke
algebras $\widehat{{ NH}}_{\pm}$ (see \cite{lucky0}, \cite{lucky}),
which contain, apart from rotation $(M_{ij})$, boost $(K_{i})$ and
space-time translation $(P_{i}, H)$ generators, the additional ones
denoted by $F_{i}$, responsible for constant acceleration. Of
course, if all generators
 $F_{i}$ are equal zero we obtain the Newton-Hooke  algebras ${NH}_{\pm}$ \cite{bacry}
 (see also \cite{contra1}-\cite{contra3}),
while for time parameter $\tau$ running to infinity we get the
acceleration-enlarged Galilei group $\,\widehat{{\mathcal G}}$
proposed in \cite{lucky1}.

In this article we discuss the role which can be played by the
acceleration-enlarged Newton-Hooke symmetries in a context of
noncommutative geometry. The suggestion to use noncommutative
coordinates goes back to Heisenberg and was formalized by Snyder in
\cite{snyder}. Recently, there were also found formal arguments
based mainly on Quantum Gravity \cite{grav1}, \cite{grav2} and
String Theory models \cite{recent}, \cite{string1} indicating that
space-time at Planck scale  should be noncommutative, i.e. it should
have a quantum nature.

Recently,  the Abelian (Reshetikhin) twist deformations (see
\cite{twist1}-\cite{twist3}) of the (ordinary\footnote{By "ordinary"
we mean the Newton-Hooke algebra without additional $F_i$
generators.}) Newton-Hooke Hopf algebras $\,{\mathcal U}_0({
NH}_{\pm})$\footnote{The Newton-Hooke Hopf algebras $\,{\mathcal
U}_0({ NH}_{\pm})$  are given by algebraic commutation relations for
${ NH}_{\pm}$ groups, supplemented by the trivial coproduct sector
$\Delta_0(a) = a\otimes 1+ 1\otimes a$.} have been proposed in
\cite{twistnh}. It has been shown that the corresponding quantum
space-times are periodic or expanding in time for $\,{\mathcal
U}_0({ NH}_{-})$ or $\,{\mathcal U}_0({ NH}_{+})$ algebras
respectively. Besides, it was also demonstrated that for
cosmological time parameter $\tau$ approaching infinity, we get the
twisted Galilei quantum groups and the corresponding canonically,
Lie-algebraically and quadratically deformed nonrelativistic
space-times \cite{dasz1}, \cite{dasz2}.

In this article we consider ten Abelian  twist deformations  of the
acceleration-enlarged Newton-Hooke Hopf algebras $\,{\mathcal
U}_0(\widehat{{ NH}}_{\pm})$. In such a way we investigate the
impact of the cosmological time $\tau$ as well as the impact of the
additional generators $F_i$ on the structure of quantum space.
Particulary, we demonstrate that due to the presence of parameter
$\tau$, the corresponding space-times can be periodic or expanding
in time for $\,{\mathcal U}_0(\widehat{{ NH}}_{-})$ or $\,{\mathcal
U}_0(\widehat{{ NH}}_{+})$ Hopf algebras respectively. Moreover, we
also provide twisted acceleration-enlarged Galilei quantum groups
and the corresponding space-times, as the $\tau \to \infty$
 limit of the considered acceleration-enlarged Newton-Hooke
Hopf structures. Surprisingly, the obtained in such a way
acceleration-enlarged Galilei quantum spaces provide (due to the
presence of generators $F_i$) the  new cubic and quartic type of
space-time noncommutativity, i.e. they take the form\footnote{$x_0 =
ct.$}
\begin{equation}
[\;{ x}_{\mu},{ x}_{\nu}\;] = i\alpha_{\mu\nu}^{\rho_1...\rho_n}{
x}_{\rho_1}...{ x}_{\rho_n}\;, \label{noncomm}
\end{equation}
with $n=3$ and $4$ respectively\footnote{There was considered in the
literature only canonical $(n=0)$, Lie-algebraic $(n=1)$ and
quadratic $(n=2)$ type of space-time noncommutativity.}.

It should be noted that two kinds of such obtained space-times
appear to be quite interesting. First of them (see formula
(\ref{gali1})) provides the deformation parameter $\beta$ with
dimension $\left[\;\beta\;\right] = \left[\;{\rm acceleration}
\times {\rm acceleration} \;\right]$, while the second one (see
formula (\ref{gali4})) provides the  parameter $\beta'$ with
dimension
 $\left[\;\beta'\;\right] = \left[\;{\rm acceleration}
\;\right]$. Such a result indicates that there can appear  a direct
link between noncommutativity and the intensively studied in the
last
 time, so-called MOND model \cite{mond}, which assumes that there
 exist in nature  observer independent acceleration
 parameter $a_0$\footnote{MOND model in a simple way explains the movies of galactic's arms at long distance scale.
 However, the proper modification of Newton equation as well as the acceleration parameter $a_0$ are introduced into model
 without any principal (theoretical) rules.}.
 Consequently, it looks  senseable to consider the simple
 classical (Newtonian) models associated with the above
 acceleration-enlarged quantum space-times. In the case of ordinary
 twisted Galilei symmetry such investigations have been performed in
\cite{walczyk1} and \cite{oscylator} respectively.

Finally, it should be mentioned that considered
acceleration-enlarged Newton-Hooke Hopf algebras and the
corresponding quantum space-times play a special role. By the proper
contractions limit $(\tau \to \infty$ or/and $F_i \to 0)$ of such
structures one can derive (or reproduce) the quantum spaces
associated with: the twisted acceleration-enlarged Galilei Hopf
algebras, the twist-deformed Newton-Hooke  quantum groups
\cite{twistnh}, and twisted  Galilei Hopf algebras \cite{dasz1},
\cite{dasz2}. For this reason, the considered spaces can be treated
as   a "source" for other  Abelian twist-deformed nonrelativistic
space-times.

The paper is organized as follows. In second section ten  Abelian
classical $r$-matrices for twisted acceleration-enlarged
Newton-Hooke Hopf algebras are considered.  The corresponding ten
quantum space-times are provided in section 3, while  their $\tau
\to \infty$ contractions to the acceleration-enlarged Galilei spaces
are discussed in section 4. Finally, two  contractions leading to
the well known (ordinary) Newton-Hooke and Galilei space-times are
mentioned in section 5. The final remarks are presented in the last
section.

\section{{{Twisted acceleration-enlarged Newton-Hooke Hopf
algebras}}}

In accordance with Drinfeld  twist procedure
\cite{twist1}-\cite{twist3}, the algebraic sector of twisted
acceleration-enlarged Newton-Hooke Hopf algebra remains undeformed
\begin{eqnarray}
&&\left[\, M_{ij},M_{kl}\,\right] =i\left( \delta
_{il}\,M_{jk}-\delta _{jl}\,M_{ik}+\delta _{jk}M_{il}-\delta
_{ik}M_{jl}\right)\;\; \;, \;\;\; \left[\, H,P_i\,\right] =\pm
\frac{i}{\tau^2}K_i
 \;,  \notag \\
&~~&  \cr &&\left[\, M_{ij},K_{k}\,\right] =i\left( \delta
_{jk}\,K_i-\delta _{ik}\,K_j\right)\;\; \;, \;\;\;\left[
\,M_{ij},P_{k }\,\right] =i\left( \delta _{j k }\,P_{i }-\delta _{ik
}\,P_{j }\right) \;,\nonumber
\\
&~~&  \cr &&\left[ \,M_{ij},H\,\right] =\left[ \,K_i,K_j\,\right] =
\left[ \,K_i,P_{j }\,\right] =0\;\;\;,\;\;\;\left[ \,K_i,H\,\right]
=-iP_i\;\;\;,\;\;\;\left[ \,P_{i },P_{j }\,\right] = 0\;,\label{nnnga}\\
&~~&  \cr &&\left[\, F_i,F_j\,\right] =\left[\, F_i,P_j\,\right]
=\left[\, F_i,K_j\,\right] =0\;\; \;, \;\;\;\left[\,
M_{ij},F_{k}\,\right] =i\left( \delta _{jk}\,F_i-\delta
_{ik}\,F_j\right) \;,\nonumber\\
&~~&  \cr &&~~~~~~~~~~~~~~~~~~~~~~~~~~~~~~~~~~~~\left[\,
H,F_{i}\,\right] =2iK_i\;,\nonumber
\end{eqnarray}
while the   coproducts and antipodes  transform as follows
\begin{equation}
\Delta _{0}(a) \to \Delta _{\cdot }(a) = \mathcal{F}_{\cdot }\circ
\,\Delta _{0}(a)\,\circ \mathcal{F}_{\cdot }^{-1}\;\;\;,\;\;\;
S_{\cdot}(a) =u_{\cdot }\,S_{0}(a)\,u^{-1}_{\cdot }\;,\label{fs}
\end{equation}
with $\Delta _{0}(a) = a \otimes 1 + 1 \otimes a$, $S_0(a) = -a$ and
$u_{\cdot }=\sum f_{(1)}S_0(f_{(2)})$ (we use Sweedler's notation
$\mathcal{F}_{\cdot }=\sum f_{(1)}\otimes f_{(2)}$).   Present in
the commutation relations (\ref{nnnga})  parameter $\tau$ denotes
the characteristic for Newton-Hooke algebra cosmological time scale
(in the limit $\tau \to \infty$ we get the acceleration-enlarged
Galilei Hopf structure $\,{\mathcal U}_0(\widehat{{\mathcal G}})$).
Besides, it should be noted, that the twist factor
$\mathcal{F}_{\cdot } \in {\mathcal U}_{\cdot}(\widehat{{
NH}}_{\pm}) \otimes {\mathcal U}_{\cdot}(\widehat{{ NH}}_{\pm})$
satisfies  the classical cocycle condition
\begin{equation}
{\mathcal F}_{{\cdot }12} \cdot(\Delta_{0} \otimes 1) ~{\cal
F}_{\cdot } = {\mathcal F}_{{\cdot }23} \cdot(1\otimes \Delta_{0})
~{\mathcal F}_{{\cdot }}\;, \label{cocyclef}
\end{equation}
and the normalization condition
\begin{equation}
(\epsilon \otimes 1)~{\cal F}_{{\cdot }} = (1 \otimes
\epsilon)~{\cal F}_{{\cdot }} = 1\;, \label{normalizationhh}
\end{equation}
with ${\cal F}_{{\cdot }12} = {\cal F}_{{\cdot }}\otimes 1$ and
${\cal F}_{{\cdot }23} = 1 \otimes {\cal F}_{{\cdot }}$.

It is well known, that the twisted algebra ${\mathcal
U}_{\cdot}(\widehat{{ NH}}_{\pm})$ can be described in terms of
so-called classical $r$-matrix $r\in {\mathcal U}_{\cdot}(\widehat{{
NH}}_{\pm}) \otimes  {\mathcal U}_{\cdot}(\widehat{{ NH}}_{\pm})$,
which satisfies the  classical Yang-Baxter equation (CYBE)
\begin{equation}
[[\;r_{\cdot},r_{\cdot}\;] ] = [\;r_{\cdot 12},r_{\cdot13} +
r_{\cdot 23}\;] + [\;r_{\cdot 13}, r_{\cdot 23}\;] = 0\;,
\label{cybe}
\end{equation}
where   symbol $[[\;\cdot,\cdot\;]]$ denotes the Schouten bracket
and for $r = \sum_{i}a_i\otimes b_i$
$$r_{ 12} = \sum_{i}a_i\otimes b_i\otimes 1\;\;,\;\;r_{ 13} = \sum_{i}a_i\otimes 1\otimes b_i\;\;,\;\;
r_{ 23} = \sum_{i}1\otimes a_i\otimes b_i\;.$$

In this article  we consider ten
 Abelian twist-deformations of acceleration-enlarged Newton-Hooke Hopf
algebra, described by the following $r$-matrices\footnote{$a\wedge b
= a\otimes b - b\otimes a$.}
\begin{eqnarray}
1)\;\;\;\;r_{\beta_1} &=&  \frac{1}{2}{\beta_1^{kl}} F_k \wedge
F_l\;\;\;\;\;\;\;\, [\;\beta_1^{kl} = -\beta_1^{lk}\;]\;,
\label{macierze01}\\&~~&\cr
 2)\;\;\;\;r_{\beta_2} &=&
\frac{1}{2}\beta_2^{kl} F_k \wedge P_l\;\;\;\;\;\;\; [\;\beta_2^{kl}
= -\beta_2^{lk}\;]\;,\label{macierze100}\\&~~&\cr
3)\;\;\;\;r_{\beta_3} &=& \frac{1}{2}{\beta_3^{kl}} K_k \wedge
F_l\;\;\;\;\;\;\, [\;\beta_3^{kl} = -\beta_3^{lk}\;]\;,
\label{macierze200}\\&~~&\cr
 4)\;\;\;\;r_{\beta_4} &=&  \beta_4 F_m
\wedge M_{kl}\;\;\; [\;m,k,l - {\rm fixed},\;\;m \neq
k,l\;]\;,\label{macierzen00}\\&~~&\cr
 5)\;\;\;\;r_{\beta_5} &=&
\frac{1}{2}{\beta_5^{kl}} P_k \wedge P_l\;\;\;\;\;\;\;\,
[\;\beta_5^{kl} = -\beta_5^{lk}\;]\;, \label{macierze0}\\
&~~&\cr 6)\;\;\;\;r_{\beta_6} &=& \frac{1}{2}\beta_6^{kl} K_k \wedge
P_l\;\;\;\;\;\;\; [\;\beta_6^{kl} = -\beta_6^{lk}\;]\;,\\ &~~&\cr
7)\;\;\;\;r_{\beta_7} &=& \frac{1}{2}{\beta_7^{kl}} K_k \wedge
K_l\;\;\;\;\;\;\, [\;\beta_7^{kl} = -\beta_7^{lk}\;]\;,
\label{macierze}\\
&~~&\cr 8)\;\;\;\; r_{\beta_8} &=&  \beta_8 K_m \wedge M_{kl}\;\;\;
[\;m,k,l - {\rm fixed},\;\;m \neq k,l\;]\;,\\ &~~&\cr 9)\;\;\;\;
r_{\beta_9} &=& \beta_9 P_m \wedge M_{kl}\;\;\; [\;m,k,l - {\rm
fixed},\;\;m \neq
k,l\;]\;,\label{macierze1}\\
&~~&\cr 10)\;\;\;\;r_{\beta_{10}} &=& \beta_{10} M_{ij} \wedge
H\;.\label{macierzenn}
\end{eqnarray}
Due to  Abelian character of the above carriers (all of them arise
from the mutually commuting elements of the algebra), the
corresponding twist factors can be get in a  standard way
\cite{twist1}-\cite{twist3}, i.e. they take the form
\begin{eqnarray}
{\cal F}_{{\beta_k }} = \exp
\left(ir_{\beta_k}\right)\;\;\;;\;\;\;k=1,2,...,10\;.
\label{factors}
\end{eqnarray}
Let us  note that   first four matrices include acceleration
generators $F_i$, while the next six  factors are the same as in the
case of Galilei and ordinary Newton-Hooke Hopf algebra, considered
in \cite{dasz1} and \cite{twistnh} respectively. Of course, for all
deformation parameters $\beta_i$ approaching zero the discussed
above Hopf structures $\,{\mathcal U}_{\beta_i}(\widehat{{
NH}}_{\pm})$ become classical, i.e. they become undeformed.

\section{{{Quantum  acceleration-enlarged Newton-Hooke spa-ce-times}}}

Let us now turn to the deformed space-times corresponding to the
twist-deformations $1)$-$10)$ discussed in pervious section. They
are defined as the quantum representation spaces (Hopf modules) for
quantum acceleration-enlarged Newton-Hooke algebras, with action of
the deformed symmetry generators satisfying suitably deformed
Leibnitz rules \cite{bloch}, \cite{wess}, \cite{chi}.

The action of generators $M_{ij}$, $K_i$, $P_i$, $H$ and $F_i$ on a
Hopf module of functions depending on space-time coordinates
$(t,x_i)$ is given by
\begin{equation}
H\rhd f(t,\overline{x})=i{\partial_t}f(t,\overline{x})\;\;\;,\;\;\;
P_{i}\rhd f(t,\overline{x})=iC_{\pm} \left(\frac{t}{\tau}\right)
{\partial_i}f(t,\overline{x})\;, \label{a1}
\end{equation}
\begin{equation}
M_{ij}\rhd f(t,\overline{x}) =i\left( x_{i }{\partial_j} -x_{j
}{\partial_i} \right) f(t,\overline{x})\;\;\;,\;\;\; K_i\rhd
f(t,\overline{x}) =i\tau \,S_{\pm} \left(\frac{t}{\tau}\right)
{\partial_i} \,f(t,\overline{x})\;,\label{dsfa}
\end{equation}
and
\begin{equation}
F_i\rhd f(t,\overline{x})=\pm 2i\tau^2\left(C_{\pm}
\left(\frac{t}{\tau}\right) -1\right)
{\partial_i}f(t,\overline{x})\;, \label{dsf}
\end{equation}
with $C_{+} [\frac{t}{\tau}] = \cosh \left[\frac{t}{\tau}\right]$,
$C_{-} [\frac{t}{\tau}] = \cos \left[\frac{t}{\tau}\right]$, $S_{+}
[\frac{t}{\tau}] = \sinh \left[\frac{t}{\tau}\right]$, $S_{-}
[\frac{t}{\tau}] = \sin \left[\frac{t}{\tau}\right]$.\\
Moreover, the $\star$-multiplication of arbitrary two functions  is
defined as follows
\begin{equation}
f(t,\overline{x})\star_{\beta_i} g(t,\overline{x}):=
\omega\circ\left(
 \mathcal{F}_{\beta_i}^{-1}\rhd  f(t,\overline{x})\otimes g(t,\overline{x})\right)
 \;,
\label{star}
\end{equation}
where symbol  $\mathcal{F}_{\beta_i}$ denotes the  twist factor (see
(\ref{factors})) corresponding to the proper acceleration-enlarged
Newton-Hooke Hopf algebra and $\omega\circ\left( a\otimes b\right) =
a\cdot b$.

In such a way  we get ten quantum   space-times
\begin{eqnarray}
&1)&[\,t,x_a\,]_{{\star}_{\beta_1}} =0\;\;\;,\;\;\;
[\,x_a,x_b\,]_{{\star}_{\beta_1}}
=4i\beta_1^{kl}\tau^4\left(C_{\pm}\left(\frac{t}{\tau}\right)
-1\right)^2 (\delta_{ak}\delta_{bl} -
\delta_{al}\delta_{bk})\;,\label{spacetime1}\\&~~&  \cr
&2)&[\,t,x_a\,]_{{\star}_{\beta_2}} =0\;,\nonumber ~\\
&&[\,x_a,x_b\,]_{{\star}_{\beta_2}} =\pm i\beta_2^{kl}\tau^2
\left(C_{\pm}\left(\frac{t}{\tau}\right) -1\right)C_{\pm}
\left(\frac{t}{\tau}\right)(\delta_{ak}\delta_{bl} -
\delta_{al}\delta_{bk})\;,\label{spacetime2}\\&~~&  \cr
&3)&[\,t,x_a\,]_{{\star}_{\beta_3}} =0\;,\nonumber\\&~~&  \cr
&&[\,x_a,x_b\,]_{{\star}_{\beta_3}} =\pm i\beta_3^{kl}\tau^3
\left(C_{\pm}\left(\frac{t}{\tau}\right) -1\right)S_{\pm}
\left(\frac{t}{\tau}\right)(\delta_{ak}\delta_{bl} -
\delta_{al}\delta_{bk})\;,\label{spacetime3}\\&~~&  \cr
&4)&[\,t,x_a\,]_{{\star}_{\beta_4}} =0\;,\nonumber ~\\&~~&  \cr
&&[\,x_a,x_b\,]_{{\star}_{\beta_4}} =\pm 4i\beta_4\tau^2
\left(C_{\pm}\left(\frac{t}{\tau}\right)
-1\right)\left[\;\delta_{ma}(x_k\delta_{bl} - x_{l}\delta_{bk}) -
\delta_{mb}(x_k\delta_{al} -
x_{l}\delta_{ik})\;\right]\;\;\;\;\;\;\;\;\label{spacetime4}\\&~~&
\cr &5)&[\,t,x_a\,]_{{\star}_{\beta_5}} =0\;\;\;,\;\;\;
[\,x_a,x_b\,]_{{\star}_{\beta_5}} =i\beta_5^{kl}\,C_{\pm}^2
\left(\frac{t}{\tau}\right)(\delta_{ak}\delta_{bl} -
\delta_{al}\delta_{bk})\;,\label{spacetime5}\\&~~&  \cr
&6)&[\,t,x_a\,]_{{\star}_{\beta_6}} =0\;\;\;,\;\;\;
[\,x_a,x_b\,]_{{\star}_{\beta_6}} =i\beta_6^{kl}\tau\, C_{\pm}
\left(\frac{t}{\tau}\right)S_{\pm}
\left(\frac{t}{\tau}\right)(\delta_{ak}\delta_{bl} -
\delta_{al}\delta_{bk})\;,\label{spacetime6}\\&~~&  \cr
&7)&[\,t,x_a\,]_{{\star}_{\beta_7}} =0\;\;\;,\;\;\;
[\,x_a,x_b\,]_{{\star}_{\beta_7}} =i\beta_7^{kl}\tau^2\, S_{\pm}^2
\left(\frac{t}{\tau}\right)(\delta_{ak}\delta_{bl} -
\delta_{al}\delta_{bk})\;,\label{spacetime7}\\&~~&  \cr
&8)&[\,t,x_a\,]_{{\star}_{\beta_8}} =0\;,\nonumber ~\\&~~&  \cr
&&[\,x_a,x_b\,]_{{\star}_{\beta_8}} =2i\beta_8\tau\, S_{\pm}
\left(\frac{t}{\tau}\right)\left[\;\delta_{ma}(x_k\delta_{bl} -
x_{l}\delta_{bk}) - \delta_{mb}(x_k\delta_{al} -
x_{l}\delta_{ak})\;\right]\;,\label{spacetime8}\\&~~&  \cr
&9)&[\,t,x_a\,]_{{\star}_{\beta_9}} =0\;,\nonumber ~\\&~~&  \cr
&&[\,x_a,x_b\,]_{{\star}_{\beta_9}} =2i\beta_9\,C_{\pm}
\left(\frac{t}{\tau}\right)\left[\;\delta_{ma}(x_k\delta_{bl} -
x_{l}\delta_{bk}) - \delta_{mb}(x_k\delta_{al} -
x_{l}\delta_{ak})\;\right]\;,\label{spacetime9a}
\\&~~&  \cr
&10)&[\,t,x_a\,]_{{\star}_{\beta_{10}}} =2i\beta_{10}
\left[\;\delta_{ia}x_j - x_{i}\delta_{ja}\;\right] \;\;\;,\;\;\;
[\,x_a,x_b\,]_{{\star}_{\beta_{10}}} =0 \;,\label{spacetime9}
\end{eqnarray}
associated with  matrices $1)$-$10)$, respectively.

 Let us note that due to the form of functions
$C_{\pm} [\frac{t}{\tau}]$ and $S_{\pm} [\frac{t}{\tau}]$ the
spatial noncommutativities 1)-9) are expanding or periodic in time
respectively. Moreover, all of them introduce classical time and
quantum spatial directions. The last type of  space-time
noncommutativity provides the quantum time and classical spatial
variables. It should be also noted that spaces 1), 2), 4), 5), 7)
and 9) are invariant with respect time reflection $t \to -t$, while
space-times 1), 3),  5), 7) and 10) - with respect $\vec{x} \to
-\vec{x}$ transformation.

Of course,  for all  deformation parameters $\beta_i$ approaching
zero, the above quantum  space-times  become commutative.

\section{{{Twisted acceleration-enlarged Galilei Hopf algebras - the $\bf \tau \to \infty$ limit}}}

In this section we provide  twisted acceleration-enlarged Galilei
Hopf algebras $\,{\mathcal U}_{\beta_i}(\widehat{\mathcal{G}})$ and
 corresponding quantum space-times, as the $\tau \to \infty$
limit of Hopf structures discussed in pervious sections. In such a
limit the commutation relations (\ref{nnnga}) become
$\tau$-independent, i.e. we neglect the impact of the cosmological
time scale $\tau$ on the structure of the considered Hopf algebras.

First of all, we perform the contraction limit $\tau \to \infty$ of
the formulas (\ref{nnnga}) and
(\ref{macierze01})-(\ref{macierzenn}). Consequently, the
corresponding classical $r$-matrices remain the same as
(\ref{macierze01})-(\ref{macierzenn}), while the algebraic sector of
all considered $\,{\mathcal U}_{\beta_i}(\widehat{\mathcal{G}})$
algebras takes the form
\begin{eqnarray}
&&\left[\, M_{ij},M_{kl}\,\right] =i\left( \delta
_{il}\,M_{jk}-\delta _{jl}\,M_{ik}+\delta _{jk}M_{il}-\delta
_{ik}M_{jl}\right)\;\; \;, \;\;\; \left[\, H,P_i\,\right] =0
 \;,  \notag \\
&~~&  \cr &&\left[\, M_{ij},K_{k}\,\right] =i\left( \delta
_{jk}\,K_i-\delta _{ik}\,K_j\right)\;\; \;, \;\;\;\left[
\,M_{ij},P_{k }\,\right] =i\left( \delta _{j k }\,P_{i }-\delta _{ik
}\,P_{j }\right) \;, \label{gggali}
\\
&~~&  \cr &&\left[ \,M_{ij},H\,\right] =\left[ \,K_i,K_j\,\right] =
\left[ \,K_i,P_{j }\,\right] =0\;\;\;,\;\;\;\left[ \,K_i,H\,\right]
=-iP_i\;\;\;,\;\;\;\left[ \,P_{i },P_{j }\,\right] = 0\;,\nonumber\\
&~~&  \cr &&\left[\, F_i,F_j\,\right] =\left[\, F_i,P_j\,\right]
=\left[\, F_i,K_j\,\right] =0\;\; \;, \;\;\;\left[\,
M_{ij},F_{k}\,\right] =i\left( \delta _{jk}\,F_i-\delta
_{ik}\,F_j\right) \;,\nonumber\\
&~~&  \cr &&~~~~~~~~~~~~~~~~~~~~~~~~~~~~~~~~~~~~\left[\,
H,F_{i}\,\right] =2iK_i\;.\nonumber
\end{eqnarray}
The corresponding coproduct sectors  can be get by application of
the formulas (\ref{fs}) and (\ref{factors}).

Let us now turn to the corresponding quantum nonrelativistic
space-times. One can check (see $\tau \to \infty$ limit of the
formulas (\ref{spacetime1})-(\ref{spacetime9})) that they look as
follows\footnote{It should be noted that the commutation relations
(\ref{gali1})-(\ref{gali9}) can be also derived  with use of the
formula (\ref{star}) and differential representation of
acceleration-enlarged Galilei generators \cite{lucky1}.}
\begin{eqnarray}
&1)&[\,t,x_a\,]_{{\star}_{\beta_1}} =0\;\;\;,\;\;\;
[\,x_a,x_b\,]_{{\star}_{\beta_1}}
=i\beta_1^{kl}\,t^4\,(\delta_{ak}\delta_{bl} - \delta_{al}\delta_{bk})\;,\label{gali1}\\
&~~&\cr &2)&[\,t,x_a\,]_{{\star}_{\beta_2}} =0\;\;\;,\;\;\;
[\,x_a,x_b\,]_{{\star}_{\beta_2}}
=\frac{i}{2}\beta_2^{kl}\,t^2\,(\delta_{ak}\delta_{bl} - \delta_{al}\delta_{bk})\;,\label{gali2}\\
&~~&\cr &3)&[\,t,x_a\,]_{{\star}_{\beta_3}} =0\;\;\;,\;\;\;
[\,x_a,x_b\,]_{{\star}_{\beta_3}} =\frac{i}{2}\beta_3^{kl}\,t^3\,
(\delta_{ak}\delta_{bl} -
\delta_{al}\delta_{bk})\;,\label{gali3}\\
&~~&\cr &4)&[\,t,x_a\,]_{{\star}_{\beta_4}} =0\;,\nonumber
~\\&~~&\cr && [\,x_a,x_b\,]_{{\star}_{\beta_4}} =2i\beta_4\,t^2\,
\left[\;\delta_{ma}(x_k\delta_{bl} - x_{l}\delta_{bk}) -
\delta_{mb}(x_k\delta_{al} -
x_{l}\delta_{ak})\;\right]\;,\label{gali4}\\
&~~&\cr &5)&[\,t,x_a\,]_{{\star}_{\beta_5}} =0\;\;\;,\;\;\;
[\,x_a,x_b\,]_{{\star}_{\beta_5}}
=i\beta_5^{kl}(\delta_{ak}\delta_{bl} - \delta_{al}\delta_{bk})\;,\label{gali5}\\
&~~&\cr &6)&[\,t,x_a\,]_{{\star}_{\beta_6}} =0\;\;\;,\;\;\;
[\,x_a,x_b\,]_{{\star}_{\beta_6}}
=i\beta_6^{kl}\,t\,(\delta_{ak}\delta_{bl} - \delta_{al}\delta_{bk})\;,\label{gali6}\\
&~~&\cr &7)&[\,t,x_a\,]_{{\star}_{\beta_7}} =0\;\;\;,\;\;\;
[\,x_a,x_b\,]_{{\star}_{\beta_7}} =i\beta_7^{kl}\,t^2\,
(\delta_{ak}\delta_{bl} - \delta_{al}\delta_{bk})\;,\label{gali7}\\
&~~&\cr &8)&[\,t,x_a\,]_{{\star}_{\beta_4}} =0\;,\nonumber\\&~~&\cr
&&[\,x_a,x_b\,]_{{\star}_{\beta_4}} =2i\beta_4\,t\,
\left[\;\delta_{ma}(x_k\delta_{bl} - x_{l}\delta_{bk}) -
\delta_{mb}(x_k\delta_{al} - x_{l}\delta_{ak})\;\right]\;,\label{gali8}\\
&~~&\cr &9)&[\,t,x_a\,]_{{\star}_{\beta_9}} =0\;,\nonumber\\&~~&\cr
&&[\,x_a,x_b\,]_{{\star}_{\beta_9}}
=2i\beta_9\,\left[\;\delta_{ma}(x_k\delta_{bl} - x_{l}\delta_{bk}) -
\delta_{mb}(x_k\delta_{al} -
x_{l}\delta_{ak})\;\right]\;,\label{gali9a}\\&~~&\cr
&10)&[\,t,x_a\,]_{{\star}_{\beta_{10}}} =2i\beta_{10}
\left[\;\delta_{ia}x_j - x_{i}\delta_{ja}\;\right] \;\;\;,\;\;\;
[\,x_a,x_b\,]_{{\star}_{\beta_{10}}} =0\;,\label{gali9}
\end{eqnarray}
in the case of $\,{\mathcal U}_{\beta_1}(\widehat{{{\mathcal G}}}),
\dots, {\mathcal U}_{\beta_{10}}(\widehat{{{\mathcal G}}})$ Hopf
algebras respectively. One can easily  see, that space-time $1)$
provides the deformation parameter $\beta$ with dimension
$\left[\;\beta\;\right] = \left[\;{\rm acceleration}\right.$ $\left.
\times {\rm acceleration} \;\right]$, while the deformation  $4)$ -
with $\left[\;\beta\;\right] = \left[\;{\rm acceleration}
\;\right]$. Due to the reasons already mentioned in Introduction
both quantum spaces   appear to be quite interesting from physical
point of view.

Obviously, for all deformation parameters $\beta_i$ approaching zero
the above Hopf algebras become classical, while the corresponding
quantum space-times - commutative.

\section{{{Twisted Newton-Hooke (and Galilei) Hopf algebras - the $F_i \to 0$ (and $\bf \tau \to \infty$) limit}}}

It should be noted, that apart of provided in pervious section
contraction limit $\tau \to \infty$ of $\,{\mathcal
U}_{\beta_i}(\widehat{{ NH}}_{\pm})$ Hopf algebras, there exist two
other contractions. First of them is defined by $F_i \to 0$ limit
and leads to the twisted $\,{\mathcal U}_{\beta_5}(\widehat{{
NH}}_{\pm}), \dots ,{\mathcal U}_{\beta_{10}}(\widehat{{
NH}}_{\pm})$ Newton-Hooke Hopf algebras and corresponding quantum
space-times, introduced in paper \cite{twistnh}. The second
contraction is given by $F_i \to 0$ and $\tau \to \infty$ limit, and
provides the twist-deformed $\,{\mathcal U}_{\beta_5}({{{\mathcal
G}}}), \dots, {\mathcal U}_{\beta_{10}}({{{\mathcal G}}})$ Galilei
Hopf algebras, proposed  (together with corresponding quantum
spaces) in the articles \cite{dasz1} and \cite{dasz2}.

\section{{{Final remarks}}}

In this article we consider ten Abelian twist-deformations of
acceleration-enlarged Newton-Hooke Hopf algebras. Besides, we
demonstrate that as in the case of ordinary twist-deformed
Newton-Hooke Hopf algebra, the corresponding spaces can be  periodic
and expanding in time for $\,{\mathcal U}_{\beta_i}(\widehat{
NH}_{-})$ and $\,{\mathcal U}_{\beta_i}(\widehat{ NH}_{+})$ quantum
groups respectively. In $\tau \to \infty$  limit we also discover
new twisted acceleration-enlarged Galilei Hopf algebras and ten
quantum space-times (\ref{gali1})-(\ref{gali9}).

It should be noted that  present studies can be extended in various
ways.  First of all, one can find the dual Hopf structures
$\,{\mathcal D}_{\beta_i}(\widehat{NH}_{\pm})$ with the use of FRT
procedure \cite{frt} or by canonical quantization of the
corresponding Poisson-Lie structures \cite{poisson}. Besides, as it
was already mentioned in Introduction, one should ask about the
basic dynamical models corresponding to the acceleration-enlarged
Newton-Hooke and Galilei space-times
(\ref{spacetime1})-(\ref{spacetime9}) and
(\ref{gali1})-(\ref{gali9}).   Finally, one can also consider more
complicated (non-Abelian) twist deformations of
acceleration-enlarged Newton-Hooke Hopf algebras, i.e. one can find
the twisted coproducts, corresponding noncommutative space-times and
dual Hopf structures. Such  problems are now under consideration.

\section*{Acknowledgments}
The author would like to thank J. Lukierski for valuable discussions.\\
This paper has been financially supported by Polish Ministry of
Science and Higher Education grant NN202318534.

\end{document}